\documentclass[angeo]{copernicus}

\usepackage{array}
\usepackage{amsmath}
\usepackage{times}
\usepackage{natbib}
\usepackage[dvips]{graphicx}
\usepackage{mathptmx}
\newcommand{\E}{\vec{E}}
\newcommand{\B}{\vec{B}}
\renewcommand{\ddot}{\cdot\cdot\,}
\newcommand{\dual}{\vphantom{F}^{\ast}\!}
\newcommand{\blank}[1]{\phantom{#1}}
\newcommand{\CompConj}[1]{\bar{#1}}

\newcommand*{\Sup}[1]{^{\mathrm{#1}}}
\newcommand{\icu}{\mathbf{i}}

\renewcommand{\vec}[1]{\mathbf{{#1}}}
\renewcommand*{\Im}[1]{\mathinner{\mathrm{Im}\left[{#1}\right]}}
\renewcommand*{\Re}[1]{\mathinner{\mathrm{Re}\left[{#1}\right]}}
\newcommand*{\tensor}[1]{\ensuremath{\boldsymbol{\mathsf{#1}}}}
\renewcommand*{\nabla}{\boldsymbol{\nabla}}

\providecommand{\tabularnewline}{\\}

\title{Canonical Electromagnetic Observables for 
Systematic Characterization of Electric and Magnetic Wave Field
Data on board
Spacecraft
}
\author[1]{J.~E.~S.~Bergman}
\author[2,*]{T.~D.~Carozzi}
\affil[1]{Swedish Institute of Space Physics, 
P.O. Box 531, SE-751 21  Uppsala, Sweden}
\affil[2]{Space Science Centre, Sussex University, Falmer, 
E. Sussex, BN1 9QT, UK}
\affil[*]{Department of Physics and Astronomy, 
University of Glasgow, Glasgow G12 8QQ, Scotland, United Kingdom}
\correspondence{J.~E.~S.~Bergman (jb@irfu.se)}
\runningauthor{J.~E.~S.~Bergman and T.~D.~Carozzi}
\runningtitle{Canonical Electromagnetic Observables}

\begin{document}

\maketitle


\begin{abstract}
We present a new characterization of partially
coherent electric and magnetic wave vector fields. 
This characterization is based on
the $36$ auto/cross correlations of the $3+3$ complex 
Cartesian components
of the electric and magnetic wave fields
and is particularly suited for analyzing electromagnetic wave data
on board spacecraft.
Data from spacecraft based electromagnetic wave instruments 
are usually processed
as data arrays. These data arrays however do not have a physical 
interpretation in themselves; 
they are simply a
convenient storage format. 
In contrast, the characterization proposed here contains exactly
the same information but 
are in the form of manifestly covariant
space-time tensors. 
We call this data format the Canonical Electromagnetic Observables (CEO) since
they correspond to unique physical observables. Some of them are already known,
such as energy density, Poynting flux, stress tensor, etc, while others
should be relevant in future space research.
As an example we use
this formalism to analyze data from
a chorus emission in the mid-latitude
magnetosphere, as recorded by the STAFF-SA instrument on board the 
Cluster-II spacecraft.
\end{abstract}

\introduction
When analyzing time varying electric and magnetic vector field data from spacecraft, 
it is common 
to construct a $6\times6$ matrix from the complex vectors
$(\vec{E},c\vec{B})$, sometimes jointly called a sixtor. 
In a Cartesian coordinate system, this energy density 
matrix (in SI units) 
can be written
\begin{align}
\epsilon_0\left(
\begin{array}{cccccc}
|E_x|^2 & E_x E_y^{\star} & E_x E_z^{\star} & cE_x B_x^{\star} & cE_x B_y^{\star} & cE_x B_z^{\star} \\
E_y E_x^{\star} & |E_y|^2 & E_y E_z^{\star} & cE_y B_x^{\star} & cE_y B_y^{\star} & cE_y B_z^{\star} \\
E_z E_x^{\star} & E_z E_y^{\star} & |E_z|^2 & cE_z B_x^{\star} & cE_z B_y^{\star} & cE_z B_z^{\star} \\
cB_x  E_x^{\star} & cB_x E_y^{\star} & cB_x E_z^{\star} & |cB_x|^2 & c^2B_x B_y^{\star} & c^2B_x B_z^{\star} \\
cB_y E_x^{\star} & cB_y E_y^{\star} & cB_y E_z^{\star} & c^2B_y B_x^{\star} & |cB_y|^2 & c^2B_y B_z^{\star} \\
cB_z E_x^{\star} & cB_z E_y^{\star} & cB_z E_z^{\star} & c^2B_z B_x^{\star} & c^2B_z B_y^{\star} & |cB_z|^2
\nonumber
\end{array}
\right)
\end{align}
This electromagnetic (EM) sixtor matrix has in various guises, such as 
wave-distribution functions (WDF) \citep{Storey:1974} and so on, been useful 
in the analysis of EM vector field data from spacecraft, for instance on the
Cluster and Polar missions. 
The second order coherency matrix is
important from a statistical viewpoint since it completely describes a 
wide sense stationary 
vector signal. 
From a physical point of view it is important since
energy density, EM wave polarization, and similar quantities can
be derived from its components.

The EM sixtor matrix can be seen as a generalization of the
coherency matrix in optics, 
which is usually a $2\times2$ Hermitian matrix, 
describing the transverse field.
The coherency matrix description
is convenient as a data storage format but
in practice, it is more common
to instead use the four Stokes parameters,
as they are physically more intuitive. 

One may ask what the Stokes description for the full EM field would be?
One way is to decompose the $6\times 6$ coherency matrix in terms of 
a complete basis set of unitary matrices, as per \citet{Samson:1980}. 
The problem with this approach is that such a decomposition is not unique;
there is an infinite number of unitary bases.

In this paper we introduce a unique set of parameters, analogous to the
Stokes parameters, 
but generalized to the full electric and magnetic 
wave fields. 
We call these parameters 
the Canonical Electromagnetic
Observables (CEO), 
due to their uniqueness, 
which comes from the fact that they
are irreducible under Lorentz transformations. Furthermore, they are not 
merely ``parameters'', they are proper space-time tensors.

Some examples of the CEO in vacuum are 
the EM energy 
density
$\epsilon_0(|\vec{E}|^2+|\vec{B}|^2)/2$,
and the energy flux density (proportional to the Poynting vector)
$\Re{\vec{E}\times\B\Sup{\ast}}/Z_0$, where 
$Z_0=\sqrt{\mu_0/\epsilon_0}$ is the vacuum impedance.
These quantities can be quite easily
identified from the sixtor matrix by inspection, for instance its 
trace is
the energy density.
Other CEO are not 
so easily identified. 

The procedure to calculate the fundamental CEO,
in terms of six real irreducible 
space-time 4-tensors, as well as the corresponding three-dimensional
representation will be given in Section 2. A two-dimensional
CEO representation is given in Section 3.

\subsection{Covariance in space physics}
Perhaps even more important for space borne observations: the EM
sixtor matrix is not covariant according to the requirements
of special relativity. Spacecraft are constantly moving and often spinning
observation platforms.
EM wave measurements become Doppler shifted and data must 
often be ``despun''.    
For the four $3\times 3$ sub-matrices: 
$\epsilon_0\vec{E}\otimes\E\Sup{\ast}$,
$\vec{E}\otimes\B\Sup{\ast}/Z_0$,
$\vec{B}\otimes\E\Sup{\ast}/Z_0$, and
$\vec{B}\otimes\B\Sup{\ast}/\mu_0$,
where $\otimes$ denote the outer (tensor) product,
despinning is straight forward by
applying rotation matrices $\tensor{R}$ from left and 
$\tensor{R}^{\rm{T}}$ from right, 
\emph{e.q.} 
$\vec{E}'\otimes\CompConj{\vec{E}}'
=\tensor{R}\vec{E}\otimes\CompConj{\vec{E}}\tensor{R}^{\rm{T}}$.
To rotate the full EM sixtor matrix, similar operations must be
performed four times. This is awkward and the resulting $6\times 6$ matrix
is still not covariant. 
For EM wave measurements in space plasma the last remark
can be crucial. 

A Lorentz boost is the
translation from one Lorentz frame to another one moving at 
velocity $\vec{v}$. A Lorentz boost does not
necessarily imply
relativistic speeds, which is a common misconception; and therefore it do not by itself
preclude what is typically associated with relativistic effects.
It is simply a quite general recipe
to make two different observers agree on a physical observation.  
The Lorentz boost of the EM
field vectors can be written 
$\vec{E}'=\gamma(\vec{E}+\vec{v}\times\vec{B})$ and
$\vec{B}'=\gamma(\vec{B}-\vec{v}\times\vec{E}/c^2)$,
where $\gamma=1/\sqrt{1-v^2/c^2}$.
As a matter of fact, the Lorentz boost is the essence of the well-known 
frozen-in field line
theorem\footnote{If $\vec{E}+\vec{v}\times\vec{B}=0$ in a plasma, 
the magnetic field lines change as though they are convected with
velocity $\vec{v}$, \emph{i.e.}, they are frozen to the plasma flow. This
is the frozen-in field line theorem of ideal MHD.} 
from magnetohydrodynamics (MHD); a theory which is
commonly used to model the solar wind plasma. In a plasma,
relativity comes 
into play at very a fundamental level since the electromagnetic 
(Lorentz) force
dominates the vast majority of all plasma interactions.

Another example illustrates the problem to separate time (frequency) 
and 
space (wave vector) in EM wave observations 
on board a spacecraft. Assume that we observe a wave mode which 
is described
by an angular frequency $\omega$ and wave vector $\vec{k}$.
We can write this
as a 4-vector $(\omega,c\vec{k})$. Let's make a Lorentz boost in the
$\vec{v}$ direction:
\begin{eqnarray}
\omega'&=&\gamma(\omega-\vec{k}\cdot\vec{v})
\label{eq:omega_boost}\\
c\vec{k}'&=&c\vec{k} +\left[\frac{\gamma-1}{v^2}(c\vec{k}\cdot\vec{v})
-\gamma\omega\right]\frac{\vec{v}}{c}
\label{eq:ck_boost}
\end{eqnarray}
What happens now for a stationary (DC) field structure moving with the solar wind plasma? 
We then have $\omega=0$ and
$|c\vec{k}|\neq 0$. For a satellite moving with velocity $\vec{v}$ relative to
the DC field structure, it is justified to 
set $\gamma \approx 1$ (the solar wind speed seldom reaches more than
$900$ km/s and  using this value we obtain $\gamma\approx1.0000045\gtrsim 1$); Eqs.
(\ref{eq:omega_boost}) and
(\ref{eq:ck_boost}) are then reduced to
\begin{eqnarray}
\omega'&\approx&-\vec{k}\cdot\vec{v}
\label{eq:omega_boost_DC}\\
c\vec{k}'&\approx&c\vec{k}
\label{eq:ck_boost_DC}
\end{eqnarray}
We can see that the DC field structure is not Lorentz contracted appreciably at 
this low velocity,
$\vec{k}'\approx\vec{k}$. However, there is a dramatic change in the observed
frequency, which for a head-on encounter with the structure is registered as 
$\omega'\approx k v$ rather than zero. 
The observed frequency is proportional to the dimension of the structure, 
which we take to be in the order of one wavelength, $\lambda=2\pi/k$. 
Taking $v=900$ km/s a 900 km DC field structure would now register
as 1 Hz, a 90 km structure as 10 Hz, and a 9 km structure as 100 Hz, etc. 

These
simple examples clearly show that 
a space-time (covariant) description is necessary even if $\gamma\gtrsim 1$. 
The frequency (time) and the associated wave vector (space)
can not be treated separately but must be considered together, as a space-time
4-tensor.
 
The Maxwell equations are 
inherently relativistic and
can easily be put into a covariant form using 4-tensors. 
From a theoretical point of view,  
this fact alone provides a very good argument why 
one should try to
express also the second order properties of the EM fields
using a covariant formalism.
This was recently carried out by the authors and published in a recent paper 
\cite{Carozzi&Bergman:JMP:2006}. In this paper we
introduced a complete set of space-time tensors, 
which can fully describe the second order
properties of EM waves. We call this set of tensors the 
Canonical Electromagnetic Observables
(CEO); in analogy with Wolf's analysis of the Stokes parameters \cite{Wolf54}.
We suggest that the CEO could be used as an alternative to
the EM sixtor matrix. Not only are the CEO covariant, but they are all real valued
and provide a useful decomposition of the sixtor matrix into convenient 
physical quantities,
especially in the three-dimensional (3D), 
so-called scalar-vector-tensor (SVT) classification; 
see section \ref{SVTform}. 
The CEO have all dimension energy density but
have various physical interpretations as will be discussed in what follows.

\section{Canonical Electromagnetic Observables}
The CEO set was derived from the 
complex Maxwell field strength
$F^{\alpha\beta}$.
Other possibilities, such as using the 4-potential $A^{\mu}$
or using a spinor formalism \cite{Barut80}, were considered but
discarded due to their lack
of physical content. The 4-potential is not directly measurable and it is
furthermore gauge dependent. Spinor formalism has been proved possible
to use \cite{Sundkvist:2006} but we believe the space-time tensor
formalism to be more intuitive and convenient to use.

In the quantum theory of light, observables of
an EM field are ultimately constructed from a complex field strength; see \cite{Wolf54}. 
The simplest of these observables are sesquilinear-quadratic 
(Hermitian quadratic) 
in $F^{\alpha\beta}$, \emph{i.e.}, they are
functions of the components of $F^{\alpha\beta} \CompConj{F}^{\gamma\delta}$,
which is a 4-tensor of rank four. 
Here we have chosen to denote the complex conjugate of the field strength 
with a bar over the field symbol in order not to confuse it with the
dual field strength, which we denote by a superscript star to the
left of the field symbol.
We showed that it was possible to
decompose 
$F^{\alpha\beta} \CompConj{F}^{\gamma\delta}$ 
into a unique set of tensors, the CEO, which
are real irreducible under the full Lorentz group.
We shall not repeat the derivation here but will instead discuss
the space-time (4-tensor) and three-dimensional (3-tensor) representations of the CEO. 

\subsection{Fundamental space-time representation}
\label{sec:4D}
In terms of the Maxwell field strength $F^{\alpha\beta}$, 
the CEO are organized
in the six real irreducible 4-tensors 
$C_{+},C_{-},Q^{\alpha\beta},T^{\alpha\beta},U^{\alpha\beta}$, and
$W^{\alpha\beta\gamma\delta}$. This is the fundamental space-time
representation of the CEO; 
their properties are listed in 
Table \ref{tab:4-tensor}.

\begin{table}[t]
\begin{tabular*}{\columnwidth}{@{\extracolsep{\fill}}|l|c|c|c|}
\hline
CEO & Rank &Proper $+$ & Number of\\
4-tensor & (Symmetry) &Pseudo $-$ & observables  \\
\hline
$C_+$ & 0 & $+$ & 1 \\
\hline
$C_-$ & 0 & $-$ & 1 \\
\hline
$T$ & 2(S) & $+$ & 9 \\
\hline
$U$ & 2(S) & $-$ & 9 \\
\hline
$Q$ & 2(A) & $-$ & 6 \\
\hline
$W$ & 4(M) & $+$ & 10 \\
\hline
\end{tabular*}
\label{tab:4-tensor}
\caption{CEO in space-time classification, \emph{i.e.}, 4-tensor notation:
$1+1+9+9+6+10=36$ observables.}
\end{table}

The CEO 4-tensors are defined as follows:
the two scalars are the vacuum proper- and pseudo-Lagrangians,
\begin{align}
C_{+}:= & \left(\CompConj{F}_{\alpha\beta}F^{\alpha\beta}
-\dual\CompConj{F}_{\alpha\beta}\dual F^{\alpha\beta}\right)/2\rm{,}
\label{eq:C+}\\
C_{-}:= & \left(\CompConj{F}_{\alpha\beta}\dual F^{\alpha\beta}
+\dual\CompConj{F}_{\alpha\beta}F^{\alpha\beta}\right)/2\rm{,}
\label{eq:C-}
\end{align}
respectively,
where we have used the dual of $F^{\alpha\beta}$ defined as\begin{equation}
\dual F^{\alpha\beta}:=\frac{1}{2}\epsilon^{\alpha\beta\gamma\delta}
F_{\gamma\delta}=\frac{1}{2}
\epsilon_{\blank{{\alpha}}\blank{{\beta}}\gamma\delta}^{\alpha\beta}
F^{\gamma\delta}.\label{eq:bivecDual}\end{equation}

The three second rank tensors consist of 
the two symmetric tensors
\begin{align}
T^{\alpha\beta}:= & \left(\CompConj{F}_{\blank{\alpha}\mu}^{\alpha}F^{\mu\beta}+\dual\CompConj{F}_{\blank{\alpha}\mu}^{\alpha}\dual 
F^{\mu\beta}\right)/2\rm{,}\label{eq:T-4}\\
U^{\alpha\beta}:= & \icu\left(\CompConj{F}_{\blank{\alpha}\mu}^{\alpha}\dual F^{\mu\beta}-\dual\CompConj{F}_{\blank{\alpha}\mu}^{\alpha}F^{\mu\beta}
\right)/2\rm{,}\label{eq:U-4}
\end{align}
and the antisymmetric tensor
\begin{align}
Q^{\alpha\beta}:= & \icu\left(\CompConj{F}_{\blank{\alpha}\mu}^{\alpha}F^{\mu\beta}-\dual\CompConj{F}_{\blank{\alpha}\mu}^{\alpha}\dual F^{\mu\beta}-2C_{+}\eta^{\alpha\beta}\right)/2\rm{.}\label{eq:Q-4}
\end{align}

The symmetric second rank tensor $T^{\alpha\beta}$ is the well-known EM
energy-stress tensor,
which contains the total energy, flux (Poynting vector),
and stress (Maxwell stress tensor)
densities.
The other two second rank tensors, $U^{\alpha\beta}$ and $Q^{\alpha\beta}$, respectively,
are less well-known. The symmetric  $U^{\alpha\beta}$ tensor is similar to 
$T^{\alpha\beta}$ in that it contains active energy densities but in $U^{\alpha\beta}$  
these densities are weighted and depend on the the handedness 
(spin, helicity, polarization, chirality)
of the EM field. 
Therefore, 
we have chosen to call them ``handed''  
energy densities. The anti-symmetric tensor $Q^{\alpha\beta}$ on the other hand is
very different in that it only contains reactive energy densities, which are
both total (imaginary part of the complex Poynting vector) and handed.

The fourth rank tensor is
\begin{eqnarray}
W^{\alpha\beta\gamma\delta}&:=& \left(\CompConj{F}^{\alpha\beta}F^{\gamma\delta}-\dual\CompConj{F}^{\alpha\beta}\dual F^{\gamma\delta}\right)/2
-2\icu Q^{[\alpha[\delta}\eta^{\gamma]\beta]}\nonumber\\
&-&\frac{2}{3}C_{+}\eta^{\alpha[\delta}\eta^{\gamma]\beta}
-\frac{1}{3}C_{-}\epsilon^{\alpha\beta\gamma\delta}
\label{eq:W}
\end{eqnarray}
where the square brackets denotes antisymmetrization over the enclosed
indices, \emph{e.g.}, $T^{\alpha[\delta}g^{\gamma]\beta}=\frac{1}{2}\left(T^{\alpha\delta}g^{\gamma\beta}-T^{\alpha\gamma}g^{\delta\beta}\right)$,
and nested brackets are not operated on by enclosing brackets, \emph{e.g.},
$T^{[\alpha[\delta}g^{\gamma]\beta]}=\frac{1}{4}\left(T^{\alpha\delta}g^{\gamma\beta}-T^{\alpha\gamma}g^{\delta\beta}-T^{\beta\delta}g^{\gamma\alpha}+T^{\beta\gamma}g^{\delta\alpha}\right)$.
It fulfills the symmetries $W^{\alpha\beta\gamma\delta}
=W^{\beta\alpha\gamma\delta}
=W^{\alpha\beta\delta\gamma}
=W^{\gamma\delta\alpha\beta}$ and $W^{\alpha[\beta\gamma\delta]}=0$.

This real irreducible rank four tensor, Eq. (\ref{eq:W}), was discovered by 
us\footnote{To the best of our knowledge, the $W^{\alpha\beta\gamma\delta}$ tensor
has never before been published in the literature.} and published in
\cite{Carozzi&Bergman:JMP:2006}, and is still under investigation; 
it is an extremely interesting geometrical object, having a structure identical
to the Weyl tensor in general relativity; see \cite{Weinberg1972}.
We have found
that it contains  
a four-dimensional generalization
of the Stokes parameters, 
as will be demonstrated in section \ref{sec:2D}
for the two-dimensional (2D)  case.
It contains both reactive total and reactive handed
energy densities.

\subsection{Three-dimensional representation}
\label{SVTform}
\begin{table}[t]
\begin{tabular*}{\columnwidth}{@{\extracolsep{\fill}}|l|c|c|c|}
\hline
& Scalars & Vectors & Tensors\\
\hline
Active Total (AT) & $u$  $\;\{S_{AT}\}$  & $\vec{P}$ $\;\{\vec{V}_{AT}\}$& 
${\tensor{T}}$ $\;\{{\tensor{T}}_{AT}\}$\\
\hline
Active Handed (AH) & $w$  $\;\{S_{AH}\}$& $\vec{V}$ $\;\{\vec{V}_{AH}\}$& ${\tensor{U}}$ $\;\{{\tensor{T}}_{AH}\}$\\
\hline
Reactive Total (RT) & $l$  $\;\{S_{RT}\}$ & $\vec{R}$ $\;\{\vec{V}_{RT}\}$ & 
${\tensor{X}}$ $\;\{{\tensor{T}}_{RT}\}$\\
\hline
Reactive Handed (RH) & $a$ $\;\{S_{RH}\}$ & $\vec{O}$ $\;\{\vec{V}_{RH}\}$ & ${\tensor{Y}}$ $\;\{{\tensor{T}}_{RH}\}$\\
\hline
\end{tabular*}
\label{tab:3-tensor}
\caption{CEO in scalar-vector-tensor (SVT) classification, 
\emph{i.e.}, 3-tensor notation:
$4\times(1+3+5)=36$ observables. In curly brackets we provide an alternative notation of the CEO.}
\end{table}
The fundamental space-time 4-tensor CEO can be written in terms of the 
three-dimensional
$\mathbf{E}$ and $\mathbf{B}$ vectors, \emph{i.e.},
3-tensors. This is convenient because it allows us to use intuitive physical  
quantities.
To systematize the 3D representation of the CEO, 
we will use a physical classification
where we organize the CEO into four groups,
which have been introduced briefly in the previous section:
the \emph{active total}, \emph{active handed}, 
\emph{reactive total}, and \emph{reactive handed} 
CEO parameter groups, respectively.  
In addition, we will use a coordinate-free 3D formalism and  
classify the CEO parameters according to rank, \emph{i.e.}, 
as
scalars, 3-vectors, and rank two 
3-tensors
(SVT classification). The 3D CEO are listed in Table \ref{tab:3-tensor}.
The alternative SVT nomenclature, enclosed by curly brackets in the
the table, was proposed in 
\cite{Olofsson:Thesis:2001},
which used 
$S$, $\vec{V}$ and
$\tensor{T}$, to denote all 
the scalars, vectors, and tensors, respectively. The 
subscript indexing scheme used here is different to that used by Olofsson. 
The CEO 3-tensors are defined as follows. 

The ``total'' parameters are:
\begin{align}
u=&\epsilon_0T^{00}=  
\epsilon_0\left(|\mathbf{E}|^{2}+|\mathbf{B}|^{2}\right)/2\\
\vec{P}=&\epsilon_0T^{i0}=  
\Re{\E\times\B\Sup{\ast}}/Z_0 \\
\tensor{T}=&\epsilon_0T^{ij}= u\mathbf{1}_{3} 
-\epsilon_0\Re{\E\otimes\E\Sup{\ast}+c^2\B\otimes\B\Sup{\ast}}\\ 
\end{align}
where $\mathbf{1}_{3}$ is the identity matrix in three dimensions. 
This
is the 3D representation of the well-known energy-stress 4-tensor 
$\epsilon_0T^{\alpha\beta}$, defined by Eq. (\ref{eq:T-4}).

The ``handed'' parameters are:
\begin{align}
w=&\epsilon_0U^{00}=  
\Im{\E\cdot\B\Sup{\ast}}/Z_0 \\
\vec{V}=&\epsilon_0U^{i0}=  -\epsilon_0
\Im{\left(\E\times\E\Sup{\ast}+c^2\B\times\B\Sup{\ast}\right)}/2\\
{\tensor{U}}=&\epsilon_0U^{ij}= v\mathbf{1}_{3} 
-\Im{ \E\otimes\B\Sup{\ast}-\B\otimes\E\Sup{\ast}}/Z_0 
\end{align}
This
is the 3D representation of the handed energy-stress 4-tensor 
$\epsilon_0U^{\alpha\beta}$, defined by Eq. (\ref{eq:U-4}).

The ``reactive total'' parameters are:
\begin{align}
l=&\epsilon_0C_{+}=  \epsilon_0
\left(|\mathbf{E}|^{2}-|\mathbf{B}|^{2}\right)/2\\
\vec{R}=&\epsilon_0Q^{i0}= 
-\Im{ \E\times\B\Sup{\ast}}/Z_0 \\
{\tensor{X}}=&\epsilon_0 W^{i0j0}= \frac{1}{2}
\left(\epsilon_0\Re{\E\otimes\E\Sup{\ast}
-c^2\B\otimes\B\Sup{\ast}}-\frac{2}{3}l\vec{1}_3\right) 
\end{align}
Contrary to the active, total and handed, parameter groups above, the reactive 
total parameter group have no single corresponding 4-tensor. 
Instead it is
composed of parts from three different CEO space-time tensors: 
the vacuum proper-Lagrangian defined by Eq. (\ref{eq:C+}), 
the reactive energy flux density from
Eq. (\ref{eq:Q-4}), and the generalized Stokes parameters corresponding
to the auto-correlated
$\vec{E}$ and $\vec{B}$ fields from Eq. (\ref{eq:W}).

The ``reactive handed''  parameters are:
\begin{align}
a=&\varepsilon_0 C_{-}=  -\Re{\E\cdot\B\Sup{\ast}}/Z_0\\ 
\vec{O}=&\varepsilon_0\frac{1}{2}\epsilon^j_{kl}Q^{kl}=  
-\frac{1}{2}\Im{\left(\E
\times\E\Sup{\ast}-c^2\B\times\B\Sup{\ast}\right)}\\
{\tensor{Y}}=&\epsilon_0\frac{1}{2}\epsilon^j_{kl}W^{i0kl}= 
\frac{1}{2}\left(\Re{ \E
\otimes\B\Sup{\ast}
+\B\otimes\E\Sup{\ast}}/Z_0-\frac{2}{3}a\vec{1}_3\right) 
\end{align}
Also for this parameter group, there is no single corresponding 4-tensor.
The reactive handed group is composed of parts from three CEO 
space-time tensors: the vacuum pseudo-Lagrangian, 
defined by Eq. (\ref{eq:C-}), the reactive handed energy flux density from
Eq. (\ref{eq:Q-4}), and the generalized Stokes parameters corresponding
to the cross-correlated
$\vec{E}$ and $\vec{B}$ fields from Eq. (\ref{eq:W}).

\section{CEO in two dimensions}
\label{sec:2D}
Up until now we have assumed that all three Cartesian components of
both the electric field, $\mathbf{E}$, and the magnetic 
field, $\mathbf{B}$, are measured. One may ask what happens if some
components are not measured; can all the 36 parameters of the CEO
be retained? Of course this is not possible, some information is certainly
lost in this case, but what one can do is to construct a set of parameters
analogous to CEO in two-dimensions. As will be shown, a total of $4\times(1+1+2)=16$ CEO 2D 
parameters can be derived.

Assume that we can measure the electric field and the magnetic 
field in a plane which we can say is the $xy$-plane without loss
of generality. Let the two-dimensional (2D) fields in this plane be denoted
$\mathbf{E}_{\mathrm{2D}}:=(E_{x},E_{y})$ and $\mathbf{B}_{\mathrm{2D}}:=(B_{x},B_{y})$,
and define the scalar product between 2D vectors as 
\begin{align}
\mathbf{E}_{\mathrm{2D}}\cdot\mathbf{B}\Sup{\ast}_{\mathrm{2D}}
=E_{x}B_{x}\Sup{\ast}+E_{y}B_{y}\Sup{\ast}
\end{align}

 and the cross product as 
 \begin{align}
\mathbf{E}_{\mathrm{2D}}\times\mathbf{B}\Sup{\ast}_{\mathrm{2D}}
=E_{x}B_{y}\Sup{\ast}-E_{y}B_{x}\Sup{\ast}
\end{align}

 and the outer (tensor) product as 
\begin{align}
\mathbf{E}_{\mathrm{2D}}\otimes\mathbf{B}\Sup{\ast}_{\mathrm{2D}}
=\left(\begin{array}{cc}
E_{x}B_{x}\Sup{\ast} & E_{x}B_{y}\Sup{\ast}\\
E_{y}B_{x}\Sup{\ast} & E_{y}B_{y}\Sup{\ast}\end{array}\right)
\end{align}

We will however not need to consider all the components of the 2D
direct product since the 2-tensors we will consider are all symmetric
and traceless. Hence, we only want the parameters which correspond
to Pauli spin matrix components 
\begin{align}
{\boldsymbol{\sigma}}_{x} & =\left(\begin{array}{cc}
0 & 1\\
1 & 0\end{array}\right)\\
{\boldsymbol{\sigma}}_{z} & =\left(\begin{array}{cc}
1 & 0\\
0 & -1\end{array}\right)\end{align}
 The Pauli components can be extracted from a 2D matrix by matrix
multiplying by a Pauli spin matrix and then taking the trace ($\mathrm{Tr}$), that
is
\begin{align}
\mathrm{Tr}\left\{ \left(\mathbf{E}_{\mathrm{2D}}
\otimes\mathbf{B}\Sup{\ast}_{\mathrm{2D}}\right){\boldsymbol{\sigma}}_{x}\right\}  
&=\left(\mathbf{E}_{\mathrm{2D}}\otimes\mathbf{B}\Sup{\ast}_{\mathrm{2D}}\right)
\ddot{\boldsymbol{\sigma}}_{x}\nonumber\\
 & :=E_{x}B_{y}\Sup{\ast}+E_{y}B_{x}\Sup{\ast}\end{align}
where the double dots, $\ddot$\,, have been 
introduced as a symbol for the
double scalar product, see \cite{Lebedev03}.

We can derive a set of two-dimensional canonical electromagnetic parameters
from the full CEO by formally taking
\begin{align}
E_{z}\equiv B_{z}\equiv0
\end{align}
and discarding all the parameters that are identically zero. In this
way we obtain the following set, which we write in the 
2D formalism introduced above. 

The ``active total'' 2D parameters are: 
\begin{align}
u_{2D} & =\epsilon_0\left(\left|\mathbf{E}_{\mathrm{2D}}\right|^{2}
+c^2\left|\mathbf{B}_{\mathrm{2D}}\right|^{2}\right)/2\nonumber\\
 & =\epsilon_0\left(\left|E_{x}\right|^{2}+\left|E_{y}\right|^{2}+\left|cB_{x}\right|^{2}+\left|cB_{y}\right|^{2}\right)/2\\
P_{z} & =\Re{ \mathbf{E}_{\mathrm{2D}}
\times\B\Sup{\ast}_{\mathrm{2D}}}/Z_0 \nonumber\\
 & =\Re{E_x B\Sup{\ast}_{y}-E_{y}B\Sup{\ast}_{x}}/Z_0 \\
T_{\sigma_{z}} & =\epsilon_0\Re{ \mathbf{E}_{\mathrm{2D}}
\otimes\E\Sup{\ast}_{\mathrm{2D}}+c^2\mathbf{B}_{\mathrm{2D}}
\otimes\B\Sup{\ast}_{\mathrm{2D}}}\ddot{\boldsymbol{\sigma}}_{z}/2\nonumber\\
 & =\epsilon_0\left(\left|E_{x}\right|^{2}-\left|E_{y}\right|^{2}
 +\left|cB_{x}\right|^{2}-\left|cB_{y}\right|^{2}\right)/2\\
T_{\sigma_{x}} & =\epsilon_0\Re{ \mathbf{E}_{\mathrm{2D}}
\otimes\E\Sup{\ast}_{\mathrm{2D}}+c^2\mathbf{B}_{\mathrm{2D}}
\otimes\B\Sup{\ast}_{\mathrm{2D}}}\ddot{\boldsymbol{\sigma}}_{x}/2\nonumber\\
 & =\epsilon_0\Re{ E_{x}E\Sup{\ast}_{y}+B_{x}B\Sup{\ast}_{y}} 
 \end{align}

The ``active handed'' 2D parameters are:
\begin{align}
w_{2D} & =\Im{ \mathbf{E}_{\mathrm{2D}}
\cdot\B\Sup{\ast}_{\mathrm{2D}}
}/Z_0\nonumber \\
 & =\Im{ E_{x}B\Sup{\ast}_{x}+E_{y}B\Sup{\ast}_{y}}/Z_0 \\
V_{z} & =\epsilon_0\Im{ \mathbf{E}_{\mathrm{2D}}
\times\E\Sup{\ast}_{\mathrm{2D}}
+c^2\mathbf{B}_{\mathrm{2D}}\times\B\Sup{\ast}_{\mathrm{2D}}
}/2\nonumber \\
&=\epsilon_0\Im{ E_{x}E\Sup{\ast}_{y}+c^2B_{x}B\Sup{\ast}_{y}}/2 \\
U_{\sigma_{z}} & =\Im{ \mathbf{E}_{\mathrm{2D}}\otimes
\B\Sup{\ast}_{\mathrm{2D}}-\mathbf{B}_{\mathrm{2D}}
\otimes\E\Sup{\ast}_{\mathrm{2D}}
}\ddot{\boldsymbol{\sigma}}_{z}/2Z_0\nonumber\\
 & =\Im{ E_{x}B\Sup{\ast}_{x}-E_{y}B\Sup{\ast}_{y}}/Z_0 \\
U_{\sigma_{x}} & =\Im{ \mathbf{E}_{\mathrm{2D}}\otimes
\B\Sup{\ast}_{\mathrm{2D}}-\mathbf{B}_{\mathrm{2D}}
\otimes\E\Sup{\ast}_{\mathrm{2D}}
}\ddot{\boldsymbol{\sigma}}_{x}/2Z_0\nonumber\\
 & =\Im{ E_{x}B\Sup{\ast}_{y}+E_{y}B\Sup{\ast}_{x}}/Z_0
 \end{align}

The ``reactive total'' 2D parameters are:
\begin{align}
l_{2D} & =\epsilon_0\left(\left|\mathbf{E}_{\mathrm{2D}}\right|^{2}
-c^2\left|\mathbf{B}_{\mathrm{2D}}\right|^{2}\right)/2\nonumber\\
 & =\epsilon_0\left(\left|E_{x}\right|^{2}+\left|E_{y}\right|^{2}
 -\left|cB_{x}\right|^{2}-\left|cB_{y}\right|^{2}\right)/2\\
R_{z} & =\Im{ \mathbf{E}_{\mathrm{2D}}
\times\B\Sup{\ast}_{\mathrm{2D}}
}/Z_0\nonumber \\
 & =\Im{ E_{x}B\Sup{\ast}_{y}-E_{y}B\Sup{\ast}_{x}}/Z_0 \\
X_{\sigma_{z}} & =\epsilon_0\Re{ \mathbf{E}_{\mathrm{2D}}
\otimes\E\Sup{\ast}_{\mathrm{2D}}
-c^2\mathbf{B}_{\mathrm{2D}}\otimes\B\Sup{\ast}_{\mathrm{2D}}} 
\ddot{\boldsymbol{\sigma}}_{z}/2\nonumber\\
 & =\epsilon_0\left(\left|E_{x}\right|^{2}-\left|E_{y}\right|^{2}
 -\left|cB_{x}\right|^{2}+\left|cB_{y}\right|^{2}\right)/2\\
X_{\sigma_{x}} & =\epsilon_0\Re{ \mathbf{E}_{\mathrm{2D}}
\otimes\E\Sup{\ast}_{\mathrm{2D}}-\mathbf{B}_{\mathrm{2D}}
\otimes\B\Sup{\ast}_{\mathrm{2D}}} 
\ddot{\boldsymbol{\sigma}}_{x}/2\nonumber\\
 & =\epsilon_0\Re{ E_{x}E\Sup{\ast}_{y}
 -c^2B_{x}B\Sup{\ast}_{y}} \end{align}

The ``reactive handed'' 2D parameters are:
\begin{align}
a_{2D} & =\Re{ \mathbf{E}_{\mathrm{2D}}\cdot\B\Sup{\ast}_{\mathrm{2D}}
}/Z_0\nonumber \\
 & =\Re{ E_{x}B\Sup{\ast}_{x}+E_{y}B\Sup{\ast}_{y}}/Z_0 \\
O_{z} & =\epsilon_0\Im{ \mathbf{E}_{\mathrm{2D}}\times
\E\Sup{\ast}_{\mathrm{2D}}
-c^2\mathbf{B}_{\mathrm{2D}}\times\B\Sup{\ast}_{\mathrm{2D}}}\nonumber \\
 & =\epsilon_0\Im{ E_{x}E\Sup{\ast}_{y}-c^2B_{x}B\Sup{\ast}_{y}} \\
Y_{\sigma_{z}} & =\Re{ \mathbf{E}_{\mathrm{2D}}
\otimes\B\Sup{\ast}_{\mathrm{2D}}+\mathbf{B}_{\mathrm{2D}}
\otimes\E\Sup{\ast}_{\mathrm{2D}}} 
\ddot{\boldsymbol{\sigma}}_{z}/2Z_0\nonumber\\
 & =\Re{ E_{x}B\Sup{\ast}_{x}-E_{y}B\Sup{\ast}_{y}}/Z_0 \\
Y{}_{\sigma_{x}} & =\Re{ \mathbf{E}_{\mathrm{2D}}
\otimes\B\Sup{\ast}_{\mathrm{2D}}+\mathbf{B}_{\mathrm{2D}}
\otimes\E\Sup{\ast}_{\mathrm{2D}}} 
\ddot{\boldsymbol{\sigma}}_{x}/2Z_0\nonumber\\
 & =\Re{ E_{x}B\Sup{\ast}_{y}+E_{y}B\Sup{\ast}_{x}}/Z_0 
 \end{align}

\begin{table}[b]
\begin{tabular*}{\columnwidth}{@{\extracolsep{\fill}}|cl|}
\hline 
Symbol&
Name\tabularnewline
\hline 
\hline 
$u_{2D}$&
Total energy\tabularnewline
\hline 
$P_{z}$$ $&
Total energy flux\tabularnewline
\hline 
$T_{\sigma_{z}}$&
Total energy stress ${\boldsymbol{\sigma}}_{z}$component\tabularnewline
\hline 
$T_{\sigma_{x}}$&
Total energy stress ${\boldsymbol{\sigma}}_{x}$component\tabularnewline
\hline
\hline 
$w_{2D}$&
Handed energy\tabularnewline
\hline 
$V_{z}$&
Handed energy flux\tabularnewline
\hline 
$U_{\sigma_{z}}$&
Handed energy stress ${\boldsymbol{\sigma}}_{z}$-component\tabularnewline
\hline 
$U_{\sigma_{x}}$&
Handed energy stress ${\boldsymbol{\sigma}}_{x}$-component\tabularnewline
\hline
\hline 
$l_{2D}$&
Vacuum proper-Lagrangian\tabularnewline
\hline 
$R_{z}$&
Reactive energy flux\tabularnewline
\hline 
$X_{\sigma_{z}}$&
EM Stokes parameter Q auto-type\tabularnewline
\hline 
$X_{\sigma_{x}}$&
EM Stokes parameter U auto-type\tabularnewline
\hline
\hline 
$a_{2D}$&
Vacuum pseudo-Lagrangian\tabularnewline
\hline 
$O_{z}$&
Reactive handed energy flux\tabularnewline
\hline 
$Y_{\sigma_{z}}$&
EM Stokes parameter Q cross-type\tabularnewline
\hline 
$Y{}_{\sigma_{x}}$&
EM Stokes parameter U cross-type\tabularnewline
\hline
\end{tabular*}
\caption{\label{CEO2Dnames}Naming scheme for the 2D CEO
parameters. }
\end{table}
\begin{figure*}
\includegraphics[width=1\textwidth,keepaspectratio]{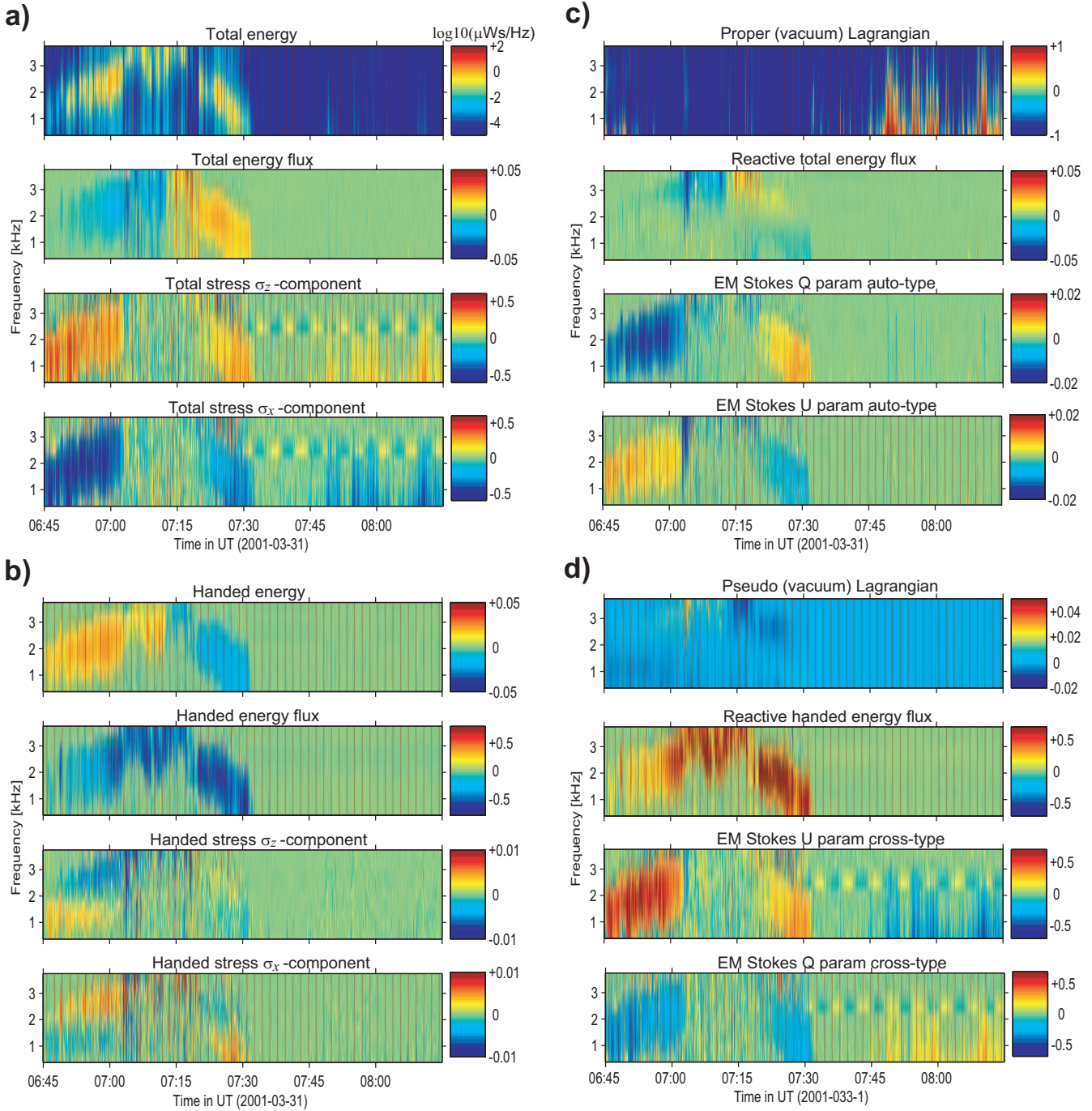}
\caption{\label{fig:Parrot03B3_4x4}Example dynamic spectra of the 16 normalized
two-dimensional CEO parameters. The parameters were computed from
STAFF-SA data from Cluster space-craft 2 using the ISDAT database
system. The following normalization has been applied: each parameter
has been divided by the total energy except the total energy itself.
Thus all spectral values are in dimensionless unit except for the
total energy. This Figure can be compared with Fig. 1 in \cite{Parrot03a}.
The 16 parameters are subdivided into a) the active total energy parameters,
b) the active handed energy parameters, c) the reactive total energy parameters,
and d) the reactive handed energy parameters.
Note that all the quantities are purely electromagnetic
in origin and so do not refer to contributions from the plasma.}
\label{fig:2D-CEO}
\end{figure*}
We can associate names with these parameters 
as listed in Table \ref{CEO2Dnames}.
The first four parameters, which we call the {}``total'' 2D CEO
parameters are all well known. 
These parameters are also known by different
names, e.g., the total energy flux is also known as the Poynting vector
($z$-component), and the total energy stress is known as the Maxwell
stress tensor (difference of diagonal components and off-diagonal
component). The remaining three sets of 2D CEO parameters are less well known.
We will not be able to provide a full physical interpretation of each
of these parameters; indeed their role in space plasma physics is
yet to be fully explored. We will only mention that the {}``handed''
parameters involve spin (helicity, chirality, polarization) 
weighted energy, i.e., the energy
of the right-hand wave modes are weighted positively and the energy
of left-hand wave modes are weighted negatively, and these weighted
energies are then added. Its flux corresponds to the concept of ellipticity
and for the case of vacuum, it is numerically equivalent to Stokes
$V$ parameter. The reactive energy densities come in two groups: 
the ``reactive total''
and the ``reactive handed''  2D CEO parameter groups. From the ``reactive total'' group,
we now recognize the reactive energy flux density, as well as the EM Stokes 
$Q$ and $U$ parameters,
which here are of the auto-type; the vacuum proper-Lagrangian needs no further 
introduction. The ``reactive handed'' group contain the handed counterparts of the
reactive energy flux density and EM Stokes parameters, which here are
of the cross-typer; the vacuum pseudo-Lagrangian is well-known.

\section{Application of CEO to Cluster data}
Let us demonstrate that the CEO parameters can easily be computed
from actual data. Assuming that we have measurements from a vector
magnetometer and an electric field instrument, all that is required
is to auto/cross-correlate all measured components and then form the
appropriate linear combination introduced above.

As an example we will consider the STAFF-SA dataset on the Cluster-II
space-craft mission \cite{Escoubet97}. 
The STAFF-SA instrument \cite{Cornilleau-Wehrlin1997} is well suited for
the CEO parameters since it outputs auto/cross-correlation of electric
and magnetic field components; however as Cluster does not measure
one of the electric field components (namely the component normal
to the spin-plane of the space-craft) we can only use the 2D
version of the CEO introduced in the previous section. 

For this particular example, we re-process the high-band part of STAFF-SA data from
an event discussed in \cite{Parrot03a} from 2001-03-31 UT. In Fig.~1 of this
paper, Parrot \emph{et al} display certain parameters based on the STAFF-SA
data computed using a numerical software package called PRASSADCO;
see \cite{Santolik2003,Santolik2006}. The interesting feature of the 2D CEO parameters is
that they are the complete set of electromagnetic field observables
in the spin-plane of the space-craft; and indeed, they use up all the
parameters in the STAFF-SA dataset expect for the magnetic field in
the spin direction. Each CEO is a distinct physical quantity and examination
of the panels in Fig. \ref{fig:Parrot03B3_4x4} indicates that this is indeed the case,
since besides showing a 
common chorus feature (the arch to the left in each panel) there are unique
points in each of the panels.

Besides being a complete description of the electromagnetic observables,
the fact that the CEO parameters are based on parameters that conform
with the physics of space-time means that we can expect physical phenomenon
to be measured properly. Seeing as how the CEO parameters have not
been explicitly measured in the past, we can expect that their future
use may lead to new physical insights, especially since several of the parameters
are completely new to space-physics. As an example consider again
the data shown in Fig \ref{fig:Parrot03B3_4x4}. It is interesting
to note that the reactive total energy flux is only significant close
to the equator; this implies that the equator is the source region
for the chorus events, since reactive energy flux is typically large
close to radiating objects due to large standing energy fields. One
can also see a modulation at $2.5$ kHz in the EM Stokes parameters.
If this is a physical phenomenon it would be indicative of Faraday
rotation. Also there seems to be frequency dispersion in the handed
stress since its components changes sign with frequency. Finally, the
handed energy clearly shows the handedness of the chorus
emissions on its own, without recourse to the sign of the total energy
flux. 

\conclusions
The proposed CEO parameters conveniently organize the measurements 
of the full EM wave field.
They are physically meaningful quantities, \emph{i.e.} they
transform as geometric (Minkowski space-time) objects and
they are mathematically unique (they are irreducible tensors).
The CEO retain all information, \emph{i.e.} nothing is lost, and 
a linear transformation back to the full sixtor form exists.
Through parameter subset selection they could
enable considerable data reduction. These parameters
have clear despinning properties and the scalar quantities 
do not even need despinning. 
Some of the CEO parameters have not been used before
to describe EM wave fields and can thus be used 
to reveal new physical insights when applied to the analysis of
EM wave field data measured by spacecraft.
To this end a particularly useful decomposition of the 36 second order 
EM components into twelve
3-tensor quantities have been provided. 
All the CEO are
real valued and unique. The active CEO can propagate to infinity.
Notably, the tensors $T^{\alpha\beta}$ and
$U^{\alpha\beta}$ obey the (vacuum) conservation laws
$\partial_\alpha T^{\alpha\beta}=0$ and 
$\partial_\alpha U^{\alpha\beta}=0$, respectively, see
\cite{Bergman&al:arxiv:2008}. The reactive CEO on the other hand do not obey
conservation laws, and hence, can not propagate to infinity. They are 
nevertheless important since they provide clues to
investigate the properties of the near field. As illustrated in Fig.\ref{fig:Parrot03B3_4x4},
this could be very usful for analysing \emph{in situ} 
measurements on board satellites when they are close to the source. 

\begin{acknowledgements}
We would like to acknowledge the financial support of the Swedish National
Space Board (SNSB) and the British Particle Physics and Astronomy Research
Council (PPARC).
\end{acknowledgements}

\bibliography{references}
\bibliographystyle{copernicus}
\end{document}